\documentclass{article}
\usepackage{dcase2022,amsmath,graphicx,url,times,booktabs, tabularx, enumitem, multirow,amssymb}
\usepackage[table,xcdraw]{xcolor}
\usepackage{array, makecell}

\def \am [#1]{\textcolor{red}{AM: #1}}

\title{Low-complexity acoustic scene classification in DCASE 2022 Challenge}
\name{Irene Mart\'{\i}n-Morat\'o$^1$, Francesco Paissan$^2$, Alberto Ancilotto$^2$, Toni Heittola$^1$,}
\secondlinename{Annamaria Mesaros$^1$, Elisabetta Farella$^2$, Alessio Brutti\thanks{This work was partially funded by the EU H2020 project MARVEL (project number: 957337).}$^2$, Tuomas Virtanen$^1$
}

 \address{$^1$Computing Sciences, Tampere University, Finland\\
\{irene.martinmorato, toni.heittola, annamaria.mesaros, tuomas.virtanen\}@tuni.fi \\
  $^2$ Fondazione Bruno Kessler, Italy\\
\{fpaissan, aancilotto, efarella, brutti\}@fbk.eu}

\begin{document}

\ninept
\maketitle

\begin{sloppy}

\begin{abstract}
This paper presents an analysis of the Low-Complexity Acoustic Scene Classification task in DCASE 2022 Challenge. The task was a continuation from the previous years, but the low-complexity requirements were changed to the following: the maximum number of allowed parameters, including the zero-valued ones, was 128 K, with parameters being represented using INT8 numerical format; and the maximum number of multiply-accumulate operations at inference time was 30 million. The provided baseline system is a convolutional neural network which employs post-training quantization of parameters, resulting in 46.5 K parameters, and 29.23 million multiply-and-accumulate operations (MMACs). Its performance on the evaluation data is 44.2\% accuracy and 1.532 log-loss. In comparison, the top system in the challenge obtained an accuracy of 59.6\% and a log loss of 1.091, having 121 K parameters and 28 MMACs. The task received 48 submissions from 19 different teams, most of which outperformed the baseline system.

\end{abstract}


\section{Introduction}
\label{sec:intro}

The task of acoustic scene classification is defined as classifying a short excerpt of audio into a class of a predefined set of classes, that indicates the context where the audio was recorded \cite{Benetos2018}. The task has been one of the main topics in the DCASE Challenge from its inception, and has developed from the original setup to include different additional problems, such as multiple devices and low-complexity conditions \cite{Heittola_2020_DCASE}. The current setup advances further towards real-world applicability by defining the low-complexity constraints, in terms of maximum number of parameters and maximum number of operations permitted at inference time, typical of current IoT devices (or microcontrollers).

For real-world applications, a classification method for acoustic scenes is expected to work in very diverse conditions, including audio captured with different devices and as short as possible inference time. The first task on low-complexity acoustic scene classification was defined in 2020 for only three classes and a single device \cite{Heittola_2020_DCASE}, for which many submissions obtained very high performance. The solutions most commonly imposed restrictions on the model architectures, using slim models and depth-wise separable CNNs. In addition, pruning and post-training quantization of the model weights were popular choices \cite{Koutini2020}.  
The data mismatch between training and testing when dealing with multiple devices has been first introduced as a separate task in 2019, and then repeated in 2020. The majority of the systems handled the mismatch through data augmentation \cite{Heittola_2020_DCASE}, with the best performance in the 2020 task being 76.5\% accuracy and 1.21 log loss \cite{Sangwon2020}.

\begin{figure}
    \centering
    \includegraphics[width=1.0\columnwidth]{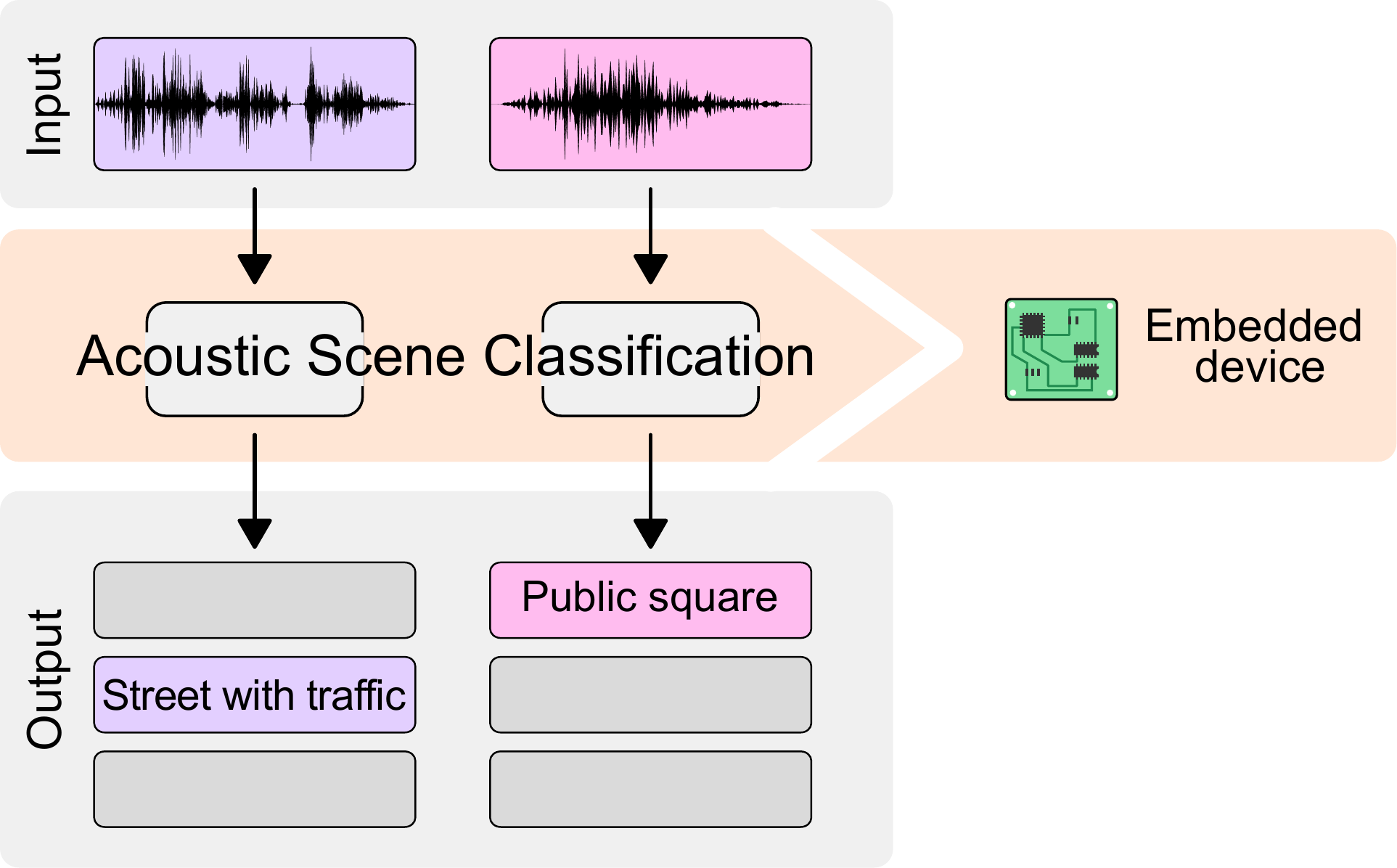}
        \caption{Acoustic scene classification for audio recordings.}
    \label{fig:overview}
    \vspace{-12pt}
\end{figure}

The combination of multiple devices and low-complexity requirements was introduced in DCASE 2021 Challenge, with the model complexity limit being set to 128 K for the non-zero parameters. This created the situation in which most high-performing systems were very close to the allowed model size limit. Sparsity used in combination with quantization emerged as a popular and efficient way of reducing the model size, while the most popular system architectures submitted in 2021 were residual models. A few teams used modified versions of available residual models suitable for processing power-constrained devices, like MobileNet \cite{Sandler2018} and EfficientNet \cite{Tan2019}. The best performing system had an accuracy of 76.1\% and log loss of 0.724 \cite{Kim2021b}, while 18 submitted systems had an accuracy above 70\%.

The current edition formulates the problem of low-complexity acoustic scene classification by defining more concretely the low-complexity limitations by selecting a class of target devices for which the developed system should be suitable. This results in the number of allowed parameters being maximum 128 K, counting all parameters, in contrast from DCASE 2021 when only the non-zero ones were counted. In addition, a limit of 30 million multiply-accumulate operations (MMACs) is approximated based on the computing power of the target device class. 

This paper introduces the results and analysis of the DCASE 2022 Challenge Task 1: Low-Complexity Acoustic Scene Classification with Multiple Devices. The paper is organized as follows: Section 2 introduces the task setup, dataset, and baseline system. Sections 3 and 4 present the challenge participation statistics and analysis of the submitted systems, respectively, while Section 5 presents conclusions and ideas for future development of this task.

\section{Task setup}
\label{sec:task-setup}

The particular aspect of this task is moving towards practical considerations that bring developed systems closer to possible target devices that impose constraints on computing power and capacity. Generalization across different devices is already a longer running feature of the acoustic scene classification task.

\subsection{Dataset and performance evaluation}

The task uses \textbf{TAU Urban Acoustic Scenes 2022}, a newly released version of the previous acoustic scene datasets. The data consists of recordings from ten acoustic scenes which represent the target classes \cite{mesaros_2019_DCASE}: \textit{airport}, \textit{indoor shopping mall}, \textit{metro station}, \textit{pedestrian street}, \textit{public square}, \textit{street with medium level of traffic}, \textit{travelling by a tram}, \textit{travelling by a bus}, \textit{travelling by an underground metro} and \textit{urban park}. Data was recorded in multiple European cities, with recordings from ten cities available in the training set and 12 in the evaluation set (two new cities compared to the training).

The audio files have been recorded simultaneously with four devices denoted A, B, C, and D, and another 11 devices denoted S1-S11 were simulated using the audio from device A. The development and evaluation sets consist of 64 and 22 hours of data, respectively. For complete details on the dataset creation and the exact amounts of data per device, we refer the reader to \cite{Heittola_2020_DCASE}. The difference from the previous datasets is that for this edition the audio data is presented in segments having a duration of 1 s, in order to comply with the inference time and computational limitations imposed by the considered target devices. 

The submissions were evaluated using multi-class cross-entropy and accuracy. Accuracy was calculated as macro-average (average of the class-wise performance for each metric), but because the data is balanced, this corresponds to the overall accuracy. The systems were ranked based on the multi-class cross-entropy (log loss), for a ranking independent of the operating point. 

As in each edition of the challenge, the audio material in the evaluation data was released two weeks prior to the challenge deadline. The participants were expected to provide class predictions for the provided audio material, and submit the system output for evaluation, together with additional information on the methods. The reference annotation of the evaluation data is only available to task organizers and was used for scoring the submissions. 

\subsection{System complexity requirements}

The computational complexity is measured in terms of parameter count and MMACs (million multiply-accumulate operations) with the requirements modeled after Cortex-M4 devices (e.g. STM32L496@80MHz or Arduino Nano 33@64MHz). 

The maximum number of parameters is 128 K, with variable type fixed into INT8, and counting all parameters. This is a major difference from DCASE 2021 in which the 128 K model size limit was only for non-zero parameters, and there was no specific format imposed on the numerical representation. This change was made because in a real operational situation, even with a sparse model, the zero-valued parameters add to the number of MACs performed at inference, and produce additional computational overhead for handling sparsity.

The maximum number of MACS per inference is 30 MMACs, approximated based on the computing power of the target device class. This limit mimics the fitting of audio buffers into SRAM (fast access internal memory) on the target device for the analysis segment of 1 s, and allows some head space for feature calculation (e.g. FFT), assuming that the most commonly used features fit under this limit.
In case learned features (embeddings) are used, e.g. VGGish~\cite{Hershey2017}, OpenL3~\cite{Cramer_2019_ICASSP} or EdgeL3~\cite{Kumari_2019_IPDPSW}, the network used to generate them contributes to the overall model size and complexity. 
Participants are required to provide full information about the model size and complexity in their technical report accompanying the submission.
To facilitate model size calculation for the challenge participant, a script for calculating the number of parameters and the MMACs is provided for Keras, TFLite and PyTorch models\footnote{https://github.com/AlbertoAncilotto/NeSsi}.

\section{Baseline system}

The baseline system has the same architecture as the 2021 one, being based on a convolutional neural network (CNN). The system consists of three CNN layers and one fully connected layer, followed by a softmax layer. The model is trained for 200 epochs with a batch size of 16. Complete details about the model and the parameters are provided with the code\footnote{https://github.com/marmoi/dcase2022\_task1\_baseline}. 
The feature extraction step follows a classical approach, where log mel-band energies are extracted every 40 ms with a 50\% hop size. This results in an input shape of $40\times51$ for each 1 second audio file. 
Post-training quantization to 8 bits is used to reduce the model complexity. The quantization was done after training, using TFLite from TensorFlow 2.0, and setting the weights to \textit{INT8} type. The baseline system has a total number of parameters of 46512. The baseline system overall performance on the development data and system complexity information are provided in Table \ref{tab:baseline}.

\begin{table}[]
\centering
\begin{tabular}{r|c|c|c}
\toprule
System &  Log loss & Accuracy & MMACs \\
\midrule
keras & 1.575 ($\pm$ 0.018) & 42.9\% ($\pm$ 0.77) & 29.23 M \\
\bottomrule
\end{tabular}
\caption{Baseline system size and performance on the development dataset. The value after $\pm$ is the standard deviation for 10 runs.}
\label{tab:baseline}
\end{table}

\section{Challenge results}
\label{sec:results}
The task received 48 submissions from a number of 19 teams. The number of participants in this edition is lower than in previous years, but similar to participation statistics of the other tasks. Only three of the 19 teams have lower performance than the baseline. The best system has a log loss of 1.091 and accuracy of 59.6\%, with the four best spots belonging to team Schmid\_CPJKU \cite{Schmid2022}.

\begin{table*}[]
    \centering
    \begin{tabular}{c|l|c|c}
    \toprule
         Rank & Label & Log loss (95\% CI) & Accuracy [\%] (95\% CI) \\
         \midrule
         1  & Schmid\_CPJKU\_3  & 1.091 (1.040 - 1.141) & 59.6 (59.4 - 59.9) \\
         5  & Chang\_HYU\_1     & 1.147 (1.081 - 1.214) &	60.8  (60.6 - 61.1) \\
         9  & Morocutti\_JKU\_task1\_4 & 1.311 (1.253 - 1.369) &	54.5  (54.2 - 54.8) \\
         11 & AI4EDGE\_IPL\_4 & 1.330 (1.281 - 1.378) &	51.6  (51.3 - 51.9) \\
         14 & Sugahara\_RION\_3	& 1.366 (1.305 - 1.426)	&51.7 (51.4 - 51.9) \\
         19 & Park\_KT\_2	& 1.431 (1.364 - 1.498)	& 52.7 (52.4 - 53.0) \\
         20 & Zou\_PKU\_1 & 1.442 (1.362 - 1.521)& 	56.3  (56.0 - 56.6)	\\
         21 & Yu\_XIAOMI\_1 & 1.456 (1.409 - 1.504)	& 46.2  (46.0 - 46.5) \\
         22 & Houyb\_XDU\_1 & 1.481 (1.416 - 1.547)	& 49.3  (49.0 - 49.5) \\
         23 & Singh\_Surrey\_3 & 1.492 (1.441 - 1.544)&	45.9  (45.6 - 46.2) \\
         \midrule
         (26) & DCASE2022 baseline & 1.532 (1.490 - 1.574) &	44.2 (44.0 - 44.5) \\
                  \bottomrule
    \end{tabular}
    \caption{Performance on the evaluation set of the best systems for top 10 teams. The first column represents the overall rank of the system among 49 (including the baseline). The baseline is not officially ranked, but its log loss corresponds in order to place 26.}
    \label{tab:results}
    \vspace{-12pt}
    \end{table*}

\subsection{Performance analysis}

The performance (log loss and accuracy) obtained by the top 10 teams, best system of each team, are presented in Table \ref{tab:results} and depicted in Figure \ref{fig:results}. The submission label was simplified to remove redundant information; submission number was kept for correspondence with the results on the website\footnote{https://dcase.community/challenge2022/task-low-complexity-acoustic-scene-classification-results}. 
The 95\% confidence intervals for log loss were calculated using the jackknife procedure.

The ranking of the systems is based on log loss, where the top ranked one is the system of Schmid\_CPJKU \cite{Schmid2022} with a log loss of 1.091. Its accuracy of 59.6\% is second-best accuracy among the 48 submissions. Team Chang\_HYU \cite{Lee2022} is ranked second by log loss, but has the overall best accuracy among all submissions. Their accuracy is 60.8\%, which appears to be significantly higher than Schmid\_CPJKU according to the 95\% confidence interval, given the amount of data in the evaluation set. 
Compared to last edition, the top accuracy has decreased by 16\%, and the log loss of the top systems is much higher. While in 2021 there were 21 systems with a log loss under 1, this year there is none. Top 10 systems have a log loss under 1.5, and an accuracy between 45.9\%-60.8\%. The decrease in performance is mostly a consequence of the data segment size being reduced from 10 to 1 second. 

\begin{figure*}
    \centering
    \includegraphics[width=1.0\textwidth]{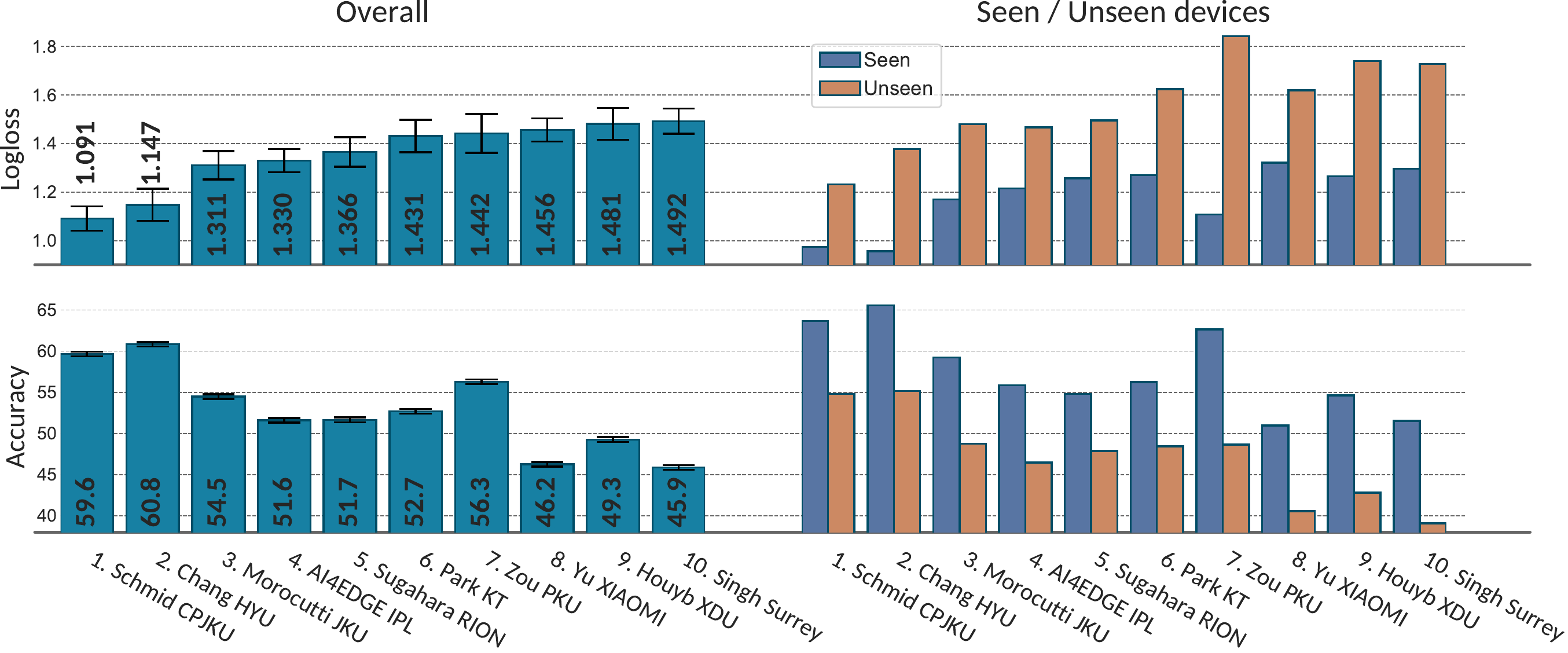}
    \caption{Classification log loss and accuracy for overall and for seen/unseen devices performance of top 10 teams, best system per team.}
    \label{fig:results}
\end{figure*}

Considering all submissions, the difference in performance between data belonging to devices seen or not in training is generally 10\% in accuracy. However, the simulated unseen devices still have a better recognition rate than data from the real device D, which is the GoPro - it appears that its characteristics are very different from those of handheld devices developed for audio (mobile phones and tablets).
Among the seen devices, the mobile devices have similar recognition rate, whether real (B, C) or simulated (S); systems have slightly better performance on device A. Given that most data in the development set belongs to device A, the relatively small difference in performance among devices shows that the systems have very strong generalization properties which cover the device mismatch. 
We also observe good generalization between seen and unseen cities, with almost no difference in classification performance between them.
Class-wise performance indicates that some acoustic scenes are more difficult overall: while scenes like bus or park obtain accuracies over 70-80\% for many systems, the large majority of systems classify scenes from pedestrian street and public square with around 30\% accuracy only.

\subsection{Machine learning characteristics}

Regarding feature extraction, all the systems make use of log-mel energies or mel spectogram, sometimes in combination with other features like deltas, spectral entropy/flatness, CQT or Gammatone. Augmentation techniques are used by most of the systems, only 5 teams do not report use of augmentation\footnote{The analysis of machine learning characteristics is based on self-reported information provided by the authors with the submission of the system outputs.}. The most popular technique is mixup (used by 33 systems), followed by SpecAugment and pitch shifting (used by 16 systems). Only one team, Zou\_PKU \cite{Xin2022} uses SpecAugment++, which is applied not only at the input but also at the hidden space of the neural network, to enhance also the intermediate feature representations. The system is ranked 7th based on the accuracy.

The most popular architectures are CNNs (used by 34 systems); some report use of MobileNet \cite{Sandler2018} (still convolutions, but depthwise separable) or BC-ResNet \cite{Kim2021b}. The use of residual models is reported by five teams, a significant reduction compared to the 2021 edition when residual networks were the most popular architecture. 

The top team Schmid\_CPJKU uses a teacher-student setup, where the PaSST models pretrained on AudioSet are used as teacher, and the student model is a RF-regularized CNN \cite{Koutini2021taslp}. The system is based on their previous submission's system reducing its complexity to fit the current constraints. For data augmentation they use Frequency MixStyle, mixing frequency-wise statistics to enhance device generalization.

\subsection{System complexity analysis}
\begin{figure}[!h]
    \centering
    \includegraphics[width=\columnwidth]{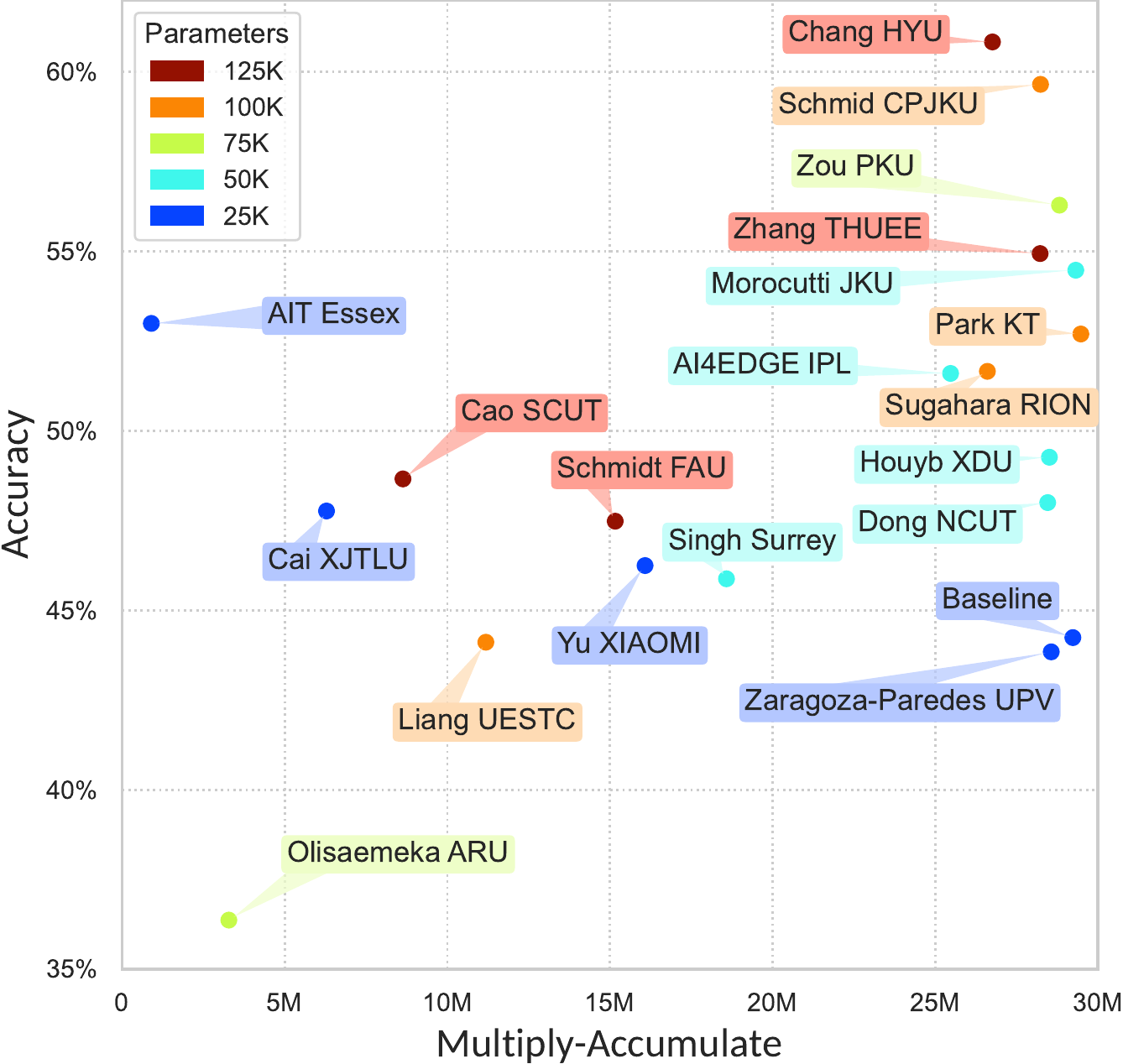}
    \caption{Analysis of performance/complexity tradeoff for different submissions.} 
    \label{acccomp}
    \vspace{-16pt}
   \end{figure}

Almost all submissions are based on inverted residual blocks, or a slight variation of this convolutional block. This is mainly because the common pattern for all participants was to adapt state-of-the-art convolutional networks to meet the computational requirements. Among the adapted networks there are MobileNets and BC-ResNets, and one submission with ShuffleNet. Other notable solutions include the use of very involved feature extraction solutions coupled with very simple neural architectures. While the networks where only slightly modified or carefully designed to meet the computational requirements (without any particular trick), a lot of focus was put on the training and data augmentation strategies. In particular, to boost inference performance, quantization-aware training (QAT) was applied by most of the participants. Another common alternative was knowledge distillation with pretrained bigger networks fine-tuned on the proposed task. Given the homogeneity in network topology of the submissions, the models proposed perform similarly in acoustic sound classification without being to diverse in computational requirements. 

Three out of the top four performing models are based on architectures characterised by large receptive fields employing, respectively, a transformer architecture, coordinate attention and an encoder-decoder architecture. This proved to optimize the performance given the limited resources available, cleverly maximizing the working memory usage of the network, as this parameter was not limited in the task description. Another notable architecture is that proposed in AIT\_Essex \cite{Pham2021}, providing almost optimal performance but very limited MMACs and/or parameter usage. This is possible thanks to their optimized convolutional block, which resembles a grouped convolution whose inputs are a combination of the original input sequence.
More standard approaches, based on BC-ResNet, inverted residual blocks or standard bi-dimensional convolutions proved less effective at solving the task with the very limited resources available. This highlights the necessity to develop and optimize neural networks specifically for different hardware platforms. In JH\_PM\_HYU \cite{Lee2022}, the authors used clever regularization techniques in order to improve the generalization capabilities of the network.
In conclusion, it is clear that, despite clever architectural designs, neural networks trained with optimized preprocessing and training strategies outperform the other approaches. In the future, it would be nice to see such techniques applied to the less computationally expensive models.
A performance versus computational cost plot containing the best performing system of each participating team is presented in Figure \ref{acccomp}.

\section{Conclusions}

This paper presented an analysis of the Low-Complexity Acoustic Scene Classification task in DCASE 2022 Challenge. The task was modeled after devices to bring the research problem closer to real-world applications. The number of multiply-and-accumulate operation set to 30 M and the total maximum number of parameters set to 128 K have been a sufficient constraint to receive a variety of interesting techniques, even though most systems were close to the imposed limits. The top systems employed large receptive fields, coordinate attention and transformer architectures to optimize performance, while quantization-aware training was the most used technique among participants to fulfill the complexity constrains. The number of submissions has decreased slightly from previous years, which may reflect the increased complexity of the task. However, the use of different devices for context-awareness is a sought after direction for applicability, therefore solutions suitable for limited computational power are needed. Moreover, the task could consider steering development towards solutions where room for improvement is still needed, like minimizing the working memory usage.

\bibliographystyle{IEEEtran}
\bibliography{refs}

\end{sloppy}
\end{document}